# CAP-DO: Learned Contextual Action Proposals for Certified Double-Oracle Solving Across Related Zero-Sum Games

Mu Wang[1*], Zhenkun Liu[1,2], Liang Liang[1,3], Guofu Zhang[1]

1 Hefei University of Technology, Hefei, 230009, Anhui, China.

2 School of Management, Nanjing University of Posts and Telecommunications, Nanjing, 210003, Jiangsu, China.

3 University of Science and Technology of China, Hefei, 230026, Anhui, China.

*Corresponding author(s). E-mail(s): wangmu@hfut.edu.cn

**Abstract**

Many security and inspection-planning problems require solving a sequence of related zero-sum games. Across this sequence, the feasible defender and attacker action spaces remain fixed, whereas each context induces a different payoff matrix through changes in target values, inspection effectiveness, costs, and interaction effects. Double Oracle (DO) solves large zero-sum games without materializing the full payoff matrix by iteratively expanding a restricted game. However, applying standard DO independently to each new payoff context requires restarting the search from a generic restricted game and requires rediscovering context-relevant actions through full-space best responses. We propose Contextual Action Proposal Double Oracle (CAP-DO), a learning-augmented framework that warm-starts DO for repeated contextual games. Offline, CAP-DO trains separate defender and attacker rankers once from solved contexts. Online, the fixed rankers propose initial restricted action sets for each new context. Learning therefore determines where certified search starts, while the current game, through full-space best-response checks and a two-sided certificate, still determines whether the output is accepted. Theoretically, CAP-DO preserves DO's full-game certification guarantee, so every accepted output meets the prescribed certificate tolerance. Under standard exact-oracle assumptions, CAP-DO also retains finite convergence when expansion is uncapped. Empirically, across three scales of a non-additive contextual inspection-game benchmark, with up to 9,880 actions per player, CAP-DO-balanced achieves higher certification rates and uses fewer full-space best-response calls than cold-start, trace-reuse, and heuristic warm starts under fixed expansion budgets. These results suggest a practical route to amortizing action discovery across related games while retaining per-instance full-space certification.

# 1 Introduction

Protecting infrastructure, transit systems, and digital services requires defenders to allocate limited inspection or auditing resources against strategic adversaries. Security-game models address this problem by computing randomized defensive policies, so that attackers cannot simply exploit fixed patrol schedules, screening rules, or auditing patterns (Ho et al. 2022; Kazeminajafabadi, Lan, and Imani 2025). In practice, the defender rarely faces a single fixed game. Threat levels, target values, inspection costs, and inspection effectiveness may change across deployments. For example, airports revise screening plans as passenger volumes and threat assessments change, and cyber defenders reallocate audits as asset values and vulnerabilities evolve. We model this as a family of contextual zero-sum games that share a combinatorial action space within each development regime but have context-dependent payoffs (Kazeminajafabadi, Lan, and Imani 2025). The computational challenge is that each new context induces a new payoff instance, so the defender must repeatedly compute and certify a randomized policy over the same large combinatorial action space.

A direct approach is to construct the full payoff matrix for each contextual game and compute an exact minimax equilibrium (Balduzzi et al. 2019; Marris et al. 2022). However, this approach becomes costly when each player selects actions in a large number of targets. Double Oracle (DO) avoids full matrix materialization by repeatedly solving a restricted game and adding full-space best responses (McMahan, Gordon, and Blum 2003). Nevertheless, standard DO is typically applied independently to each game instance. When the context changes, it is typically rerun from a generic restricted game and must rediscover useful actions online.

In repeated deployment, previously solved contexts leave certified DO traces. These traces record actions added as full-space best responses or retained in the final equilibrium support. Because the feasible action space and payoff-generating structure are shared within a regime, these traces provide context-labeled evidence about useful actions. Contexts within a regime share the same feasible action space and payoff structure. Historical traces can therefore guide a context-dependent warm start for a new game (Sambharya et al. 2024). However, payoff changes can alter strategic relevance. Historical actions are thus proposals rather than certificates. This raises a natural question of whether historical traces can reduce repeated action discovery while full-space verification preserves per-instance certification? The answer requires a separation between proposal and verification: past traces may suggest useful actions, but the current game must still be certified against its full defender and attacker action spaces.

We propose Contextual Action Proposal Double Oracle (CAP-DO), a learning-augmented DO framework for repeated contextual zero-sum games. Our contributions are threefold: First, we introduce a trace-supervised proposal mechanism that converts certified traces from previously solved contexts into reusable action proposals. The proposal model is trained once offline and then used to initialize restricted games for future contexts. Second, we prove that this learned initialization does not weaken DO's certification logic. Accepted outputs satisfy full-game saddle-gap bounds, and CAP-DO inherits finite convergence under exact best responses and unbounded expansion. Third, we evaluate CAP-DO on non-additive combinatorial inspection games across exact finite, exact-value bridge, and oracle-only large-action regimes. The results show that learned proposals reduce certified search effort and improve tail certificate behavior compared with cold-start, historical-reuse, and heuristic warm starts. Overall, CAP-DO shows how learned warm starts can accelerate repeated action discovery while certification determines whether the resulting strategy is accepted.

## 2 Related Work

DO reduces equilibrium computation in large zero-sum games by repeatedly solving a restricted game and adding full-space best responses until no profitable deviation remains (McMahan, Gordon, and Blum 2003). PSRO extends the double-oracle idea to policy spaces by iteratively training response policies and solving a meta-game over the discovered policy set (Lanctot et al. 2017). Subsequent variants improve scalability through parallel response-oracle training and extensive-form decompositions, including Pipeline PSRO and XDO/NXDO (McAleer et al. 2020; McAleer et al. 2021). Recent lower-bound results further show that DO can require many iterations in worst-case zero-sum settings, highlighting that best-response expansion itself can be a computational bottleneck (Zhang and Sandholm 2024). These methods address large strategy spaces, but they primarily target solving a given game instance or population game, rather than amortizing information across many related contextual games.

Other work integrates learning with game solving across varying formulations. Online and regret-minimizing DO variants employ no-regret updates to accelerate the equilibrium search within a single large game instance(Tang et al. 2023; Dinh et al. 2021). There are also exact sequence-form DO algorithms for two-player zero-sum extensive-form games with imperfect information and graph-based zero-sum security games (Bosansky et al. 2014; Jain et al. 2011). In dynamic settings, stochastic games model environmental transitions driven by endogenously coupled player actions over time. Separately, policy-learning and neural-equilibrium methods directly map a game instance to a policy or equilibrium estimate (Marris et al. 2022; Kamra et al. 2018). In both cases, the learned output serves as the solution estimate for the current game. Our learned model instead proposes initial restricted action sets. A certified DO solver then recomputes and verifies the mixed strategies.

A source of inspiration comes from learning-augmented optimization and algorithms with predictions, where data-driven predictions accelerate empirical performance while classical algorithms preserve classical worst-case correctness (Cohen-Addad et al. 2024). Borrowing this insight, mathematical optimization frequently leverages warm starts to expedite related solves without altering underlying optimality checks (Sambharya et al. 2024). In our setting, the warm start takes the form of context-dependent restricted action sets. They are informed by certified search information from prior contexts and do not represent a numerical iterate, policy, or equilibrium estimate. The two-sided DO certificate still verifies the resulting strategies against the full defender and attacker action spaces.

## 3 Problem Setup

We study repeated contextual zero-sum games in which many related finite games must be solved under changing operational conditions. For a fixed structural regime $(T, k, b)$, a context $x$ induces a finite loss game

$$G_x = (\mathcal{D}_T, \mathcal{A}_T, L_x),$$

where $\mathcal{D}_T$ and $\mathcal{A}_T$ are the defender and attacker pure action spaces, and $L_x(d, a)$ is the defender loss for $d \in \mathcal{D}_T$ and $a \in \mathcal{A}_T$. In the inspection setting, $\mathcal{D}_T$ contains feasible inspection plans of size $k$, while $\mathcal{A}_T$ contains attack sets of size $b$. The context $x$ is observed before solving the current game. Across contexts within the same structural regime, feasible actions remain fixed, but their payoffs change through $L_x$.

The context is multidimensional rather than a single payoff parameter. It may encode target values, inspection effectiveness, target-level costs, pairwise attack synergies, and defender route or conflict costs. Thus, repeated instances share the same combinatorial decision structure while inducing different payoff landscapes and different strategically relevant actions.

This setting creates a tension between reuse and certification. Solving each context from a cold-start forces DO to rediscover context-relevant actions online. Previously solved contexts provide traces of actions that were useful under related payoff conditions, but equilibrium support and best-response structure can shift when the context changes. Historical actions are therefore informative as proposals, but they cannot certify a strategy for a new game. The computational problem is not to predict an equilibrium directly, but to predict a small context-conditioned restricted action set that reduces certified search effort while preserving full-space certification for each new context.

### 4 Contextual Action Proposal Double Oracle

CAP-DO exploits the fact that repeated contextual games are not solved in isolation: previously solved contexts reveal which defender and attacker actions were useful under related payoff conditions. It turns these certified traces into context-conditioned action rankings and uses the top-ranked actions to initialize the restricted game for each new context. After this learned initialization, the online solver follows the certified DO loop with exact restricted solving, full-space best-response checks, expansion, and two-sided certification. Figure 1 summarizes the architecture.

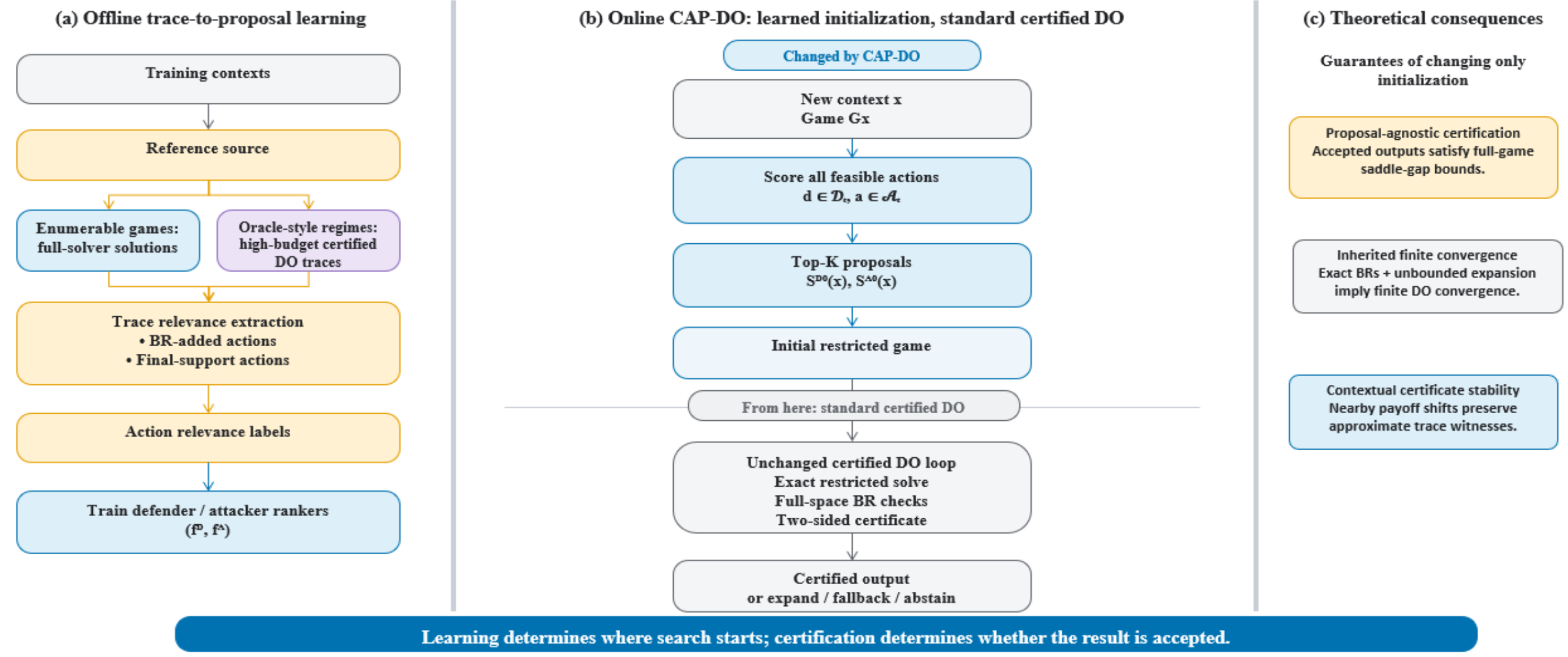

*Figure 1: Overview of CAP-DO. Offline, solved training contexts are converted into action-relevance labels using best-response additions and final-support actions, which train context-conditioned defender and attacker rankers. Online, these rankers propose top-K initial restricted action sets for a new context. After this learned initialization, CAP-DO enters the standard certified DO loop, where restricted solving, full-space best-response checks, and two-sided certification determine acceptance or further expansion.*

### 4.1 Trace-Supervised Action Proposals

For each training context $x_i$, CAP-DO obtains reference supervision from either a full solver, when the full game is enumerable, or a high-budget certified DO trace. When full-solver supervision is available, equilibrium probabilities can be used as soft action-importance labels. In the oracle-only regimes, CAP-DO instead constructs binary trace-relevance labels from certified DO traces. In both cases, the offline goal is not to learn equilibrium probabilities directly, but to identify actions that are useful for constructing a certifiable restricted game.

For a certified DO trace, CAP-DO marks an action as relevant if it satisfies at least one of two conditions: it was added by a full-space best-response expansion, or it received positive probability in the final certified restricted solution. Let $B_D(x_i)$ and $B_A(x_i)$ denote the defender and attacker actions added by best-response expansion in the reference trace. Let $p_i$ and $q_i$ be the final lifted restricted-game strategies. The relevant action sets are

$$R_D(x_i) = B_D(x_i) \cup \{d \in \mathcal{D}_T : p_i(d) > 0\},$$
$$R_A(x_i) = B_A(x_i) \cup \{a \in \mathcal{A}_T : q_i(a) > 0\}.$$

Actions in $R_D(x_i)$ and $R_A(x_i)$ receive label one, and the remaining actions receive label zero. These labels indicate trace relevance, not calibrated equilibrium probability. They identify actions that helped the reference procedure build a certifiable restricted game.

CAP-DO then trains separate defender and attacker rankers. For an action $u$ and context $x$, the rankers output scores

$$s_D(u, x) = f_D(\phi_D(u, x)), \qquad s_A(u, x) = f_A(\phi_A(u, x)),$$

where $\phi_D$ and $\phi_A$ are action-context features. The scores are used only to rank actions. They are not interpreted as mixed-strategy probabilities, and the proposal models are trained once offline and then fixed for all held-out contexts. For the inspection game, $\phi_D$(d, x) includes aggregate features of the inspected target set, including selected target values, value-effectiveness products, inspection effectiveness, inspection costs, pairwise synergy and route-conflict aggregates, top-value counts, and diversity features. The attacker feature vector $\phi_A$(a, x) includes selected target-value aggregates, pairwise attack synergy, top-value counts, and diversity features. These hand-designed features are used only for ranking feasible pure actions; they are not mixed strategies and are not interpreted as calibrated probabilities.

### 4.2 Online Initialization and Certified Expansion

For a new context $x$, CAP-DO scores all feasible defender and attacker actions and selects the top-ranked actions:

$$S_D^0(x) = \mathrm{TopK}_{d \in \mathcal{D}_T}\, s_D(d, x),$$
$$S_A^0(x) = \mathrm{TopK}_{a \in \mathcal{A}_T}\, s_A(a, x).$$

These sets form the initial restricted game. From this point onward, CAP-DO follows the standard DO loop: it solves the restricted zero-sum game exactly, computes full-space defender and attacker best responses, and either accepts the solution if the two-sided certificate passes or expands the restricted game with the exposed best responses, as shown in Technical Supplement Figure 1.

For lifted restricted-game strategies $(p, q)$, define the full-space upper and lower bounds

$$\overline{v}_x(p) = \max_{a \in \mathcal{A}_T} p^\top L_x e_a, \qquad \underline{v}_x(q) = \min_{d \in \mathcal{D}_T} e_d^\top L_x q,$$

and the certificate gap

$$\Delta_x(p,q) = \overline{v}_x(p) - \underline{v}_x(q).$$

CAP-DO accepts only if

$$\Delta_x(p,q)/s_x \le \tau,$$

where $s_x$ is a payoff-normalization scale and $\tau$ is the certificate threshold. If the certificate fails, the full-space best responses are added to the restricted game. If the expansion budget is exhausted before certification, the instance is either passed to a fallback solver when available or marked uncertified.

**Algorithm CAP-DO**

**Input:** training contexts, reference solutions or certified DO traces, new context $x$, proposal cap $K$, certificate threshold $\tau$, expansion budget $R$.

**Offline:** For each training context $x_i$, obtain full-solver labels or a high-budget certified DO trace. Convert each reference trace into defender and attacker relevance sets $R_D(x_i)$ and $R_A(x_i)$. Train defender and attacker rankers $f_D$ and $f_A$ on action-context features and relevance labels.

**Online:** Score all defender and attacker actions under the new context $x$. Initialize the restricted action sets with $S_D^0(x)$ and $S_A^0(x)$ from the top-ranked actions. For at most $R$ expansion rounds, solve the restricted zero-sum game exactly and compute full-space best-response bounds. If $\Delta_x(p,q)/s_x \le \tau$, return the lifted strategies $(p,q)$ as certified. Otherwise, add the exposed full-space best responses to the restricted game and continue. If the budget is exhausted, use a fallback solver if available; otherwise return uncertified.

### 4.3 Theoretical Properties

All statements below are conditional on finite action spaces, exact restricted-game solving, exact full-space best responses, and nonempty initial restricted sets. The lifted strategies are evaluated against the original defender and attacker action spaces, not only against the current restricted game.

The following results characterize the consequences of changing only the initialization interface. The first two properties show that learning does not weaken the certification and convergence mechanisms inherited from DO. The third property gives a sufficient condition under which trace-supported proposals can transfer across nearby contexts.

**Proposition 1. Proposal-agnostic certification.**
If CAP-DO accepts $(p,q)$ with

$$\Delta_x(p,q) \le \epsilon,$$

then $(p,q)$ is an $\epsilon$-approximate saddle point of the full game $G_x$. If CAP-DO uses the normalized rule

$$\Delta_x(p,q)/s_x \le \tau,$$

then the unnormalized saddle gap is at most $\tau s_x$.

**Proof sketch.** Let $v_x$ be the full-game value. By minimax duality,

$$\underline{v}_x(q) \le v_x \le \overline{v}_x(p).$$

Therefore, both players' possible improvements are bounded by the interval width $\Delta_x(p,q)$. This argument is independent of how the initial restricted sets were generated.

**Proposition 2. Inherited finite convergence.**
For any nonempty initial restricted sets $S_D^0 \subseteq \mathcal{D}_T$ and $S_A^0 \subseteq \mathcal{A}_T$, if restricted games are solved exactly, full-space best responses are exact, and the expansion budget is unbounded, CAP-DO terminates in finitely many iterations.
A conservative expansion bound is

$$N_{\max} \le |\mathcal{D}_T| - |S_D^0| + |\mathcal{A}_T| - |S_A^0|.$$

**Proof sketch.** CAP-DO differs from standard DO only in the initial restricted sets. Whenever the certificate fails, at least one full-space best response outside the current restricted game is added. Under exact solving, if both full-space best responses are already contained in the current restricted sets, the restricted equilibrium is also certified for the full game. Hence, whenever the certificate fails, at least one newly exposed best response can be added. Since $\mathcal{D}_T$ and $\mathcal{A}_T$ are finite, this can happen only finitely many times. Once no improving full-space best response remains, the two-sided gap is zero or below the prescribed threshold.

**Proposition 3. Contextual certificate stability.**
Let x and x' have identical action spaces. For any fixed lifted strategy pair (p, q), if $\| L_x - L_{x'} \|_{\max} \le \delta$ . Then $|\Delta_x(p,q) - \Delta_{x'}(p,q)| \le 2\delta$. In particular:
$\Delta_x(p,q) \le \Delta_{x'}(p,q) + 2\delta$.
Thus, a small certificate gap at x' carries to x with additive degradation no greater than 2δ. This statement concerns a fixed pair of strategies; it does not assert that proposal rankings transfer across contexts.
**Proof sketch.** Each one-sided full-space best-response bound changes by at most δ when every payoff entry changes by at most δ. Their difference therefore changes by at most 2δ. The general Lipschitz form and complete proof appear in the supplementary material. Details can be found in Technical Supplement S1.

**Proposition 4 (Recall-sensitive repair bound).**

Fix a deterministic certified reference DO trace for context x, with the same seed, tolerance, and tie-breaking. Let $C_{i(x)}$ be the distinct player-i best-response additions in that trace, and let $U_{i(x)}$ be the distinct actions newly added by an unbudgeted CAP-DO run after failed certificate checks. For the initial proposal $S_i^{0(x)}$, define

$$m_x = \sum_{i \in \{D,A\}} \left| C_i(x) \setminus S_i^0(x) \right|, h_x = \sum_{i \in \{D,A\}} |U_i(x) \setminus C_i(x)|$$

Here, $m_x$ is the recall deficit relative to the canonical trace, and $h_x$ counts off-reference additions.

Assume finite action spaces, monotone expansion, exact restricted solves and full-space best responses, and that each failed certificate check adds a returned action outside the current restricted game. If $E_x$ is the number of expansions, then $E_x \leq m_x + h_x$. If $B_x$ counts both player full-space best-response calls, with one initial check and one check after each expansion, then

$$B_x = 2(E_x + 1) \leq 2(1 + m_x + h_x)$$

**Proof sketch.** Every expansion contributes at least one distinct online-added action. Partitioning those actions into canonical-trace and off-reference actions yields the bound. The detailed proof clarifies why recall alone does not bound the trajectory-dependent term $h_x$. Details can be found in Technical Supplement S2.

## 5 Experiments and Results

### 5.1 Experimental Protocol

We evaluate proposal quality, certification, and search cost across three regimes, and use a no-context ablation to isolate context conditioning. CAP-DO was evaluated on a contextual non-additive inspection-game family. Each instance is a finite zero-sum loss game in which the defender inspects $k = 3$ targets and the attacker attacks $b = 3$ targets. For a regime with $T$ targets, the defender's pure actions are feasible inspection sets and the attacker's pure actions are feasible attack sets. Contexts change the target values, inspection effectiveness, target-level costs, pairwise attack synergies, and defender route or conflict costs, while the feasible action spaces remain fixed within each scale.

We evaluate CAP-DO under three complementary regimes. Table 1 summarizes their roles. The T16 regime is a finite exact benchmark. The T30 regime is the central bridge experiment because proposals are trained from certified DO traces while exact values are computed only after each method returns. The T40 regime is an oracle-only scale stress test in which the dense payoff matrix is not materialized and exact exploitability is not reported.

**Table 1. Evaluation regimes.**

| Regime | T | Actions/player | Supervision | role |
|---|---|---|---|---|
| Exact finite | 16 | 560 | Full-solver labels | Finite sanity check |
| Exact-value bridge | 30 | 4060 | Certified DO traces | Main validation |
| Oracle-only scale | 40 | 9880 | Certified DO traces | Scale stress |

For T=16, each seed uses 200 training contexts, 50 calibration contexts, and 100 test contexts, with five random seeds. For T=30 and 40, each seed uses 100 training contexts and 50 test contexts, with three random seeds, giving 150 evaluated seed-context instances per regime. T30 exact full-solver values are computed only after each method returns. T40 does not use exact full-solver evaluation. Full-space best responses are exact over the finite action list. In T=16, separation uses the materialized payoff matrix. In T=30 and 40, we enumerate all $\binom{T}{3}$ feasible defender and attacker combinations as action lists and evaluate the expected loss against the current restricted mixed strategy using vectorized submatrix computations. We do not materialize the full dense payoff matrix and do not use a separate combinatorial optimization oracle.

All cap-based warm starts use the same proposal cap $K = 50$. We compare cold-start DO, random-$K$ warm starts, frequency-$K$ trace reuse, nearest-neighbor-$K$ trace reuse, a hand-coded heuristic-$K$ warm start, CAP-DO-fast, and CAP-DO-balanced. The no-context learned ablation uses the same trace labels and the same model family as CAP-DO-balanced, but removes all current-context payoff features. Cold-start DO uses a smaller standard seed and is interpreted as the classical search baseline. The T30 and T40 learned proposals are trained from certified DO traces.
Random-K samples K defender and K attacker actions uniformly without replacement. Frequency-K ranks actions by how often they appear as relevant actions in training traces. Nearest-neighbor-K retrieves the closest training context using standardized context-level features and reuses that context's trace-induced action ranking. Heuristic-K uses a hand-coded inspection heuristic: defender actions are scored by high value-effectiveness coverage and low inspection or route-conflict cost, while attacker actions are scored by high target value and pairwise attack synergy. All cap-based warm starts use the same proposal cap K = 50 in the main experiments.

We separate evaluation metrics into four layers. First, exact-evaluation metrics report normalized exact exploitability, but only for T16 and T30, where exact post-hoc values are available. Second, certificate metrics report the normalized two-sided full-space certificate gap and certified success rate in all regimes. The normalized gap is

$$\hat{\Delta}_x(p, q) = \frac{\Delta_x(p, q)}{s_x},$$

where $s_x$ is the payoff-normalization scale and $\Delta_x(p, q)$ is the full-space certificate gap defined in the theory section. A run is certified when $\hat{\Delta}_x(p, q) \leq \tau$, with $\tau = 0.05$ in the main experiments. Third, search-effort metrics report expansions, full-space best-response calls, and fallback or abstention rates. Fourth, Runtime is reported separately as one-time offline cost and per-context online cost. Online time includes action scoring, top-K selection, restricted-game solving, full-space best responses, and certification; offline cost includes reference-trace processing, feature construction, and ranker training. For repeated solving, the offline cost is amortized across contexts as offline cost divided by the number of solved contexts. On T40, reference-trace generation costs 285.633 s per seed, while incremental learning after traces are available costs 8.605 s with Ridge and 12.909 s with HistGB. Proposal construction adds 76.807 ms or 120.799 ms per new context; the full breakdown is reported in Technical Supplement Table A.

All methods are evaluated on matched seed–context pairs. We use a paired hierarchical bootstrap that resamples training seeds first and test contexts within each seed; thus, the 150 seed–context evaluations are not treated as 150 independent training repetitions. For all metrics, we report paired estimates and differences with 95% confidence intervals; certification rates are reported as percentage-point differences.

Contexts are sampled independently within each structural regime. Target locations are fixed on a one-dimensional grid, while each context samples a latent hotspot center and width. Target values are generated from this hotspot field with noise and normalized to unit mean. Inspection effectiveness and target-level inspection costs are sampled independently. Pairwise attack synergy $W_{\{ij\}}$ and defender route/conflict cost $C_{\{ij\}}$ are distance-decayed random matrices, symmetrized and normalized over positive target pairs. Thus, train and test games share feasible action spaces within a regime but differ in values, effectiveness, costs, and pairwise interaction structure. CAP-DO-fast uses a ridge-regression ranker with standardized action-context features. CAP-DO-balanced uses histogram gradient boosting rankers with the same supervision source and the same proposal cap.

### 5.2 Main Bridge Evaluation at T30

The T30 bridge regime is the central test of CAP-DO. In this setting, proposal models are trained from certified DO traces rather than from test-context full-game solutions. Online solving uses restricted linear programs (LPs) and full-space best-response oracles, while exact values are computed only after each method returns for evaluation. This regime therefore tests whether trace-trained contextual proposals improve certified search under an online protocol that resembles the larger oracle-only setting.

As shown in Figure 2, under the tight expansion budget $R = 3$, CAP-DO-balanced gives the best combination of certification rate, tail exploitability, and best-response effort. It certifies 88.0% of the 150 matched seed–context evaluations, compared with 74.0% for the heuristic, 50.0% for nearest-neighbor trace reuse, 48.7% for frequency reuse, and 0.7% for cold-start DO. It also has the lowest p95 normalized exact exploitability, 0.024, compared with 0.039 for the heuristic and 0.058 for frequency. At the same time, it reduces mean full-space BR calls to 3.773, compared with 5.827 for the heuristic, 6.973 for frequency, and 8.000 for cold-start DO.

Under the larger budget R = 10, weaker warm starts partially catch up in certified success because DO has more opportunities to repair poor initial restricted games. However, CAP-DO-balanced still retains the best tail-risk/search-effort profile. It certifies all 150 matched seed-context instances (100.0%), achieves a p95 normalized exact exploitability of 0.021 (95% hierarchical bootstrap CI: 0.012–0.026), and uses 4.187 mean full-space BR calls (95% CI: 3.667–4.747). The strongest non-learned p95 exploitability is nearest-neighbor at 0.024, while the heuristic uses 6.747 BR calls.

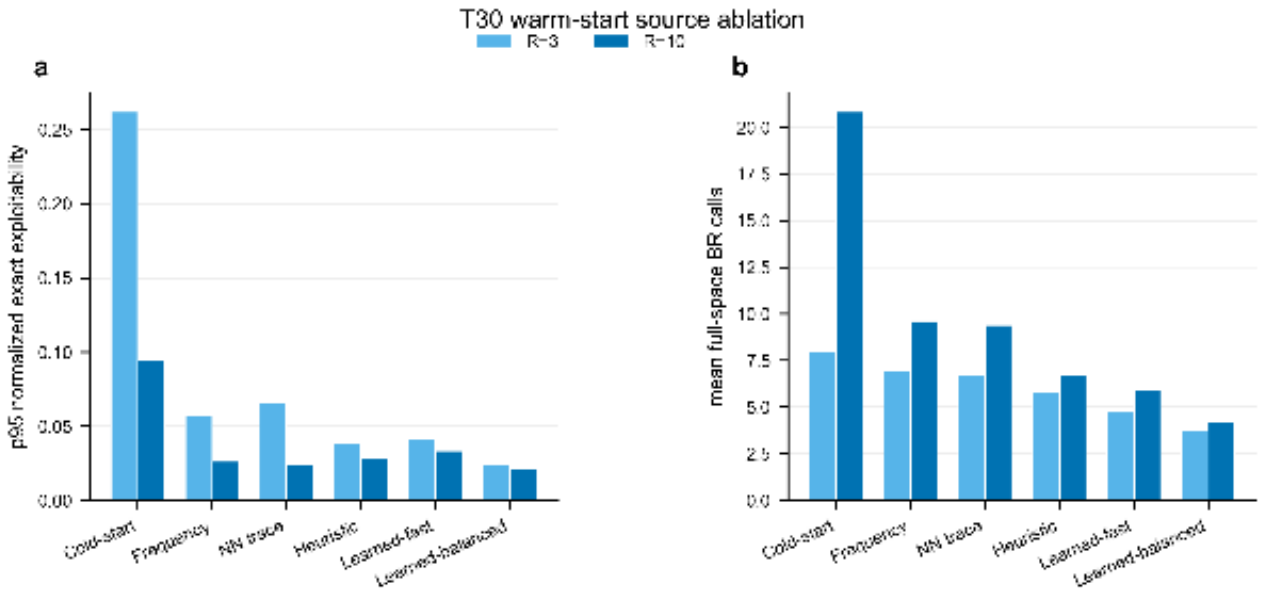

*Figure 2. T30 bridge warm-start comparison. Panel (a) reports p95 normalized exact exploitability, and panel (b) reports mean full-space best-response (BR) calls under $R = 3$ and $R = 10$. Lower values are better. Learned-balanced achieves the lowest tail exploitability and uses the fewest BR calls, especially under the tight budget $R = 3$.*

### 5.3 Oracle-Only Scale Stress at T40

The T40 regime evaluates CAP-DO at a larger action scale, with 9,880 actions per player. In this regime, the dense payoff matrix is not materialized and no full-game solver is called. Accordingly, T40 evaluates certification rate, certificate gap, and oracle effort rather than exact exploitability. As shown in Figure 3, CAP-DO-balanced gives the strongest certified-search profile at T40. Under the tight budget R=3, CAP-DO-balanced certifies 90.7% of the matched seed-context instances (95% CI: 84.0–96.0%) and achieves a p95 normalized certificate gap of 0.085 (95% CI: 0.048–0.387), compared with 52.7% certified success and a 0.205 p95 gap for the heuristic warm start. Under R=10, CAP-DO-balanced reaches 100.0% certification while using 4.413 mean full-space BR calls (95% CI: 3.733–5.187), fewer than learned-fast (7.013), heuristic (9.347), frequency (13.960), and nearest-neighbor (13.947). Since exact exploitability is unavailable at T40, the evidence supports CAP-DO's ability to improve certification and reduce oracle effort in the oracle-only regime.

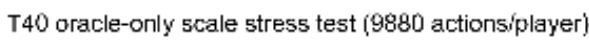

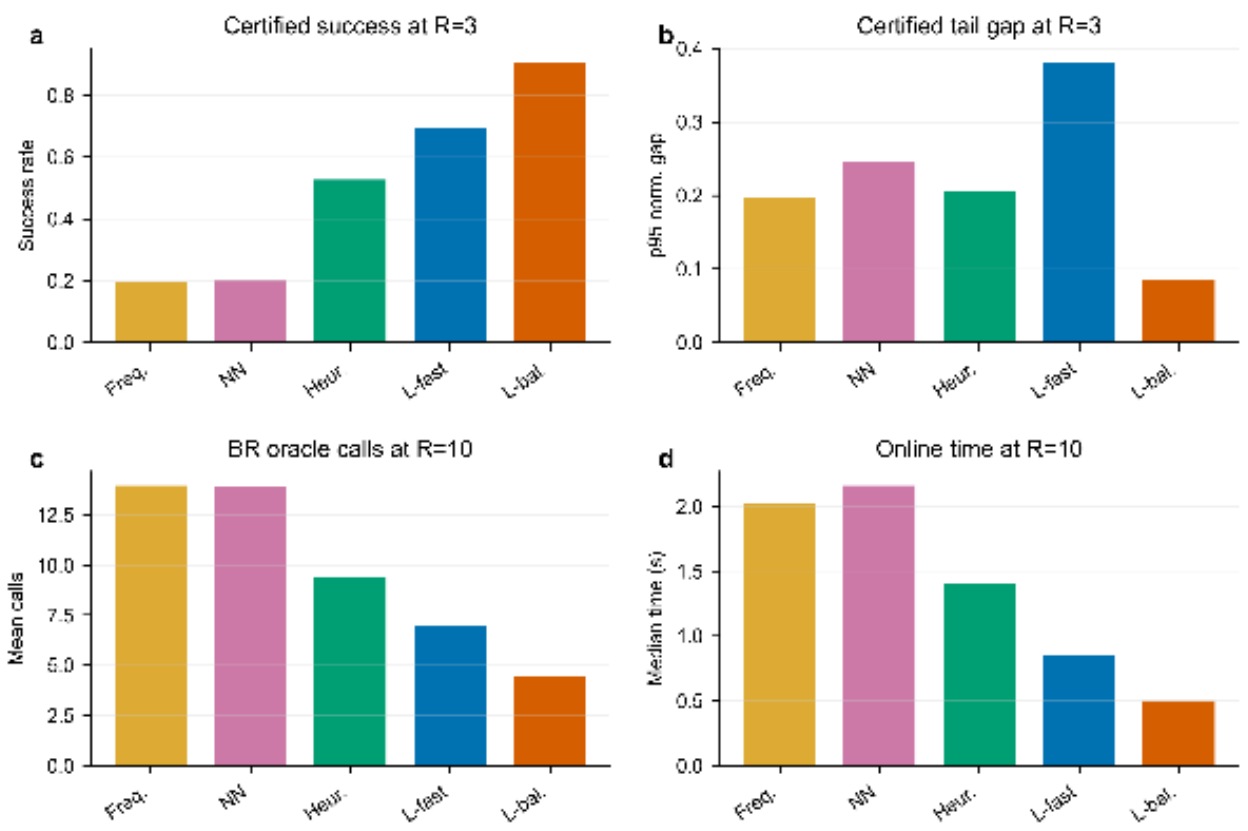

Figure 3. *T40 oracle-only scale stress. Exact exploitability is not reported because the full game is not materialized. Panels a and b show certified success and p95 normalized certificate gap under R = 3. Panels c and d show mean full-space BR calls and median online time under R = 10. CAP-DO-balanced has the highest R = 3 certification rate 90.7%, the lowest R = 3 p95 certificate gap 0.085, and the fewest R = 10 BR calls 4.413.*

### 5.4 Context Conditioning (ablations)

At fixed K=50, we next isolate the effect of current-context features. Larger warm-start size is controlled by using the same proposal cap for all cap-based baselines. To isolate the role of context conditioning, we include a no-context learned ablation that uses the same trace-derived labels and the same model family as CAP-DO-balanced, but removes all current-context payoff features. It can therefore learn globally useful or frequently relevant actions, but it cannot adapt its ranking to the observed payoff context. We therefore use the T30 test set to examine proposal coverage, initial certificate quality, and paired online-repair differences at a fixed proposal size of K = 50.

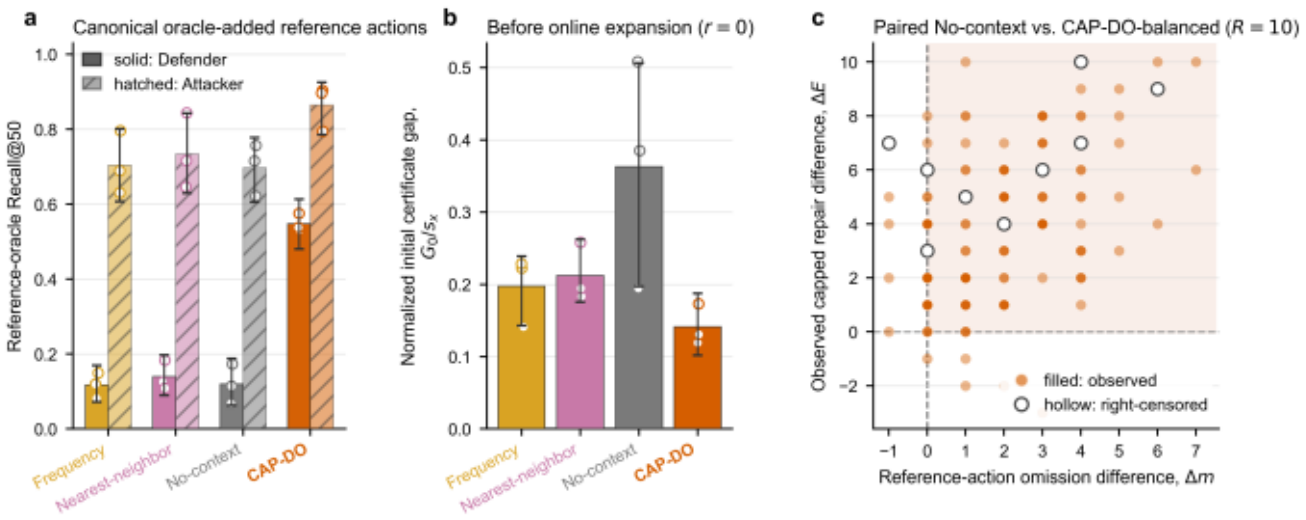

Figure 4. *Mechanism analysis on the T30 bridge test set with K = 50. (a) Recall of oracle-added actions from a canonical reference DO trace; solid and hatched bars denote defender and attacker actions. (b) Normalized initial certificate gap $G_0/s_x$, where $G_0 = \Delta_x(p^0, q^0)$, before online expansion ( $r = 0$ ). (c) Paired No-context–CAP-DO differences, $\Delta m = m_{NoCtx} - m_{CAP}$ and $\Delta E = min(E_{NoCtx}, 10) - min(E_{CAP}, 10)$, the upper-right quadrant favors CAP-DO-balanced. Bars in (a)–(b) show means with paired 95% confidence intervals. Hollow markers in (c) indicate right-censored pairs at R = 10; the panel contains 150 matched seed–context evaluations. Reference traces are used only for post-hoc analysis.*

Figure 4 provides mechanism-consistent evidence for the gains of CAP-DO-balanced. At the same proposal size, CAP-DO-balanced achieves the highest recall of oracle-added reference actions, with the largest improvement on the defender side (Fig. 4a). It also has the lowest point estimate of the normalized initial certificate gap before expansion (Fig. 4b). In the paired comparison, 97 of 150 seed–context evaluations lie in the upper-right quadrant, indicating fewer reference-action omissions and fewer observed capped repairs for CAP-DO-balanced (Fig. 4c). The mean observed differences are Δm = 1.807 and ΔE = 3.933. Because repair counts are capped at R = 10 and reference traces are not unique completion paths, these results support the proposed mechanism but do not imply that recall deterministically determines the subsequent DO trajectory. Supporting checks and sensitivity analyses can be found in Technical Supplement.

## 6 Discussion

CAP-DO shows that learning can be introduced into repeated zero-sum game solving without replacing certified equilibrium computation. Its central contribution is to move learning to the initialization interface of DO. Instead of predicting an equilibrium directly, CAP-DO converts certified traces from previously solved contexts into reusable action proposals. These proposal rankers are trained once offline and then used to initialize many future contextual games. This shifts part of the repeated action-discovery burden away from online certified search, while leaving restricted solving, full-space best-response checks, expansion, and two-sided certification unchanged. This amortization is most beneficial for repeated contextual solving. For a one-shot game, the offline learning cost may not be recovered.

Once the proposal rankers are trained, future contexts can start from more informative restricted games instead of rediscovering useful actions from a generic initialization. Our experiments support that learned proposals reduce certified search effort and improve tail certificate behavior, especially when the online expansion budget is tight. CAP-DO assigns different roles to learning and certification: learned proposals provide starting points for search, while certification remains tied to the current game.

**Technical Supplement S1 Detailed Proof of Proposition 3**

For contexts x and x' with identical action spaces, use the entrywise maximum norm for payoff perturbations. The one-sided bounds and certificate gap are as defined in the main paper.

For any fixed lifted strategy pair (p,q), the condition

$$\| L_x - L_{x'} \|_{\max} \leq \delta$$

implies

$$|\Delta_x(p,q) - \Delta_{x'}(p,q)| \leq 2\delta$$

**Proof.**

For fixed p, every pure attacker response changes the expected loss by at most δ; taking a maximum preserve this bound. Likewise, for fixed q, the lower best-response bound changes by at most δ. Subtracting the two bounds gives the displayed 2δ result.

**Lipschitz corollary.**

If the contextual payoff map satisfies

$$\| L_x - L_{x'} \|_{\max} \leq C_{\mathcal{L}} \| x - x' \|_{\mathcal{X}}$$

then

$$|\Delta_x(p,q) - \Delta_{x'}(p,q)| \leq 2C_{\mathcal{L}} \| x - x' \|_{\mathcal{X}}$$

Limitation. This is a fixed-strategy stability result, not a trace- or ranking-transfer theorem; CAP-DO still requires current-context full-space certification.

**Technical Supplement S2 Detailed Proof of Proposition 4**

Use the same deterministic trace, C_i(x), U_i(x), S_i^0(x), m_x, and h_x as in the main-paper statement. The backslash in the following displays denotes set difference.

**Proposition S2.** Recall-sensitive repair bound.

Under finite action spaces, monotone action-set expansion, exact restricted solves, exact full-space best responses, and the stated new-action condition, the number of expansions satisfies

$$E_x \leq m_x + h_x$$

When B_x counts both player best-response calls, using one initial check and one check after each expansion, the corresponding call bound is

$$B_x = 2(E_x + 1) \leq 2(1 + m_x + h_x)$$

**Proof.**

No action newly added online can have been present in the initial restricted action set. Therefore, every online-added action that also belongs to the canonical trace lies among the canonical actions omitted by the proposal:

$$U_i(x) \cap C_i(x) \subseteq C_i(x) \setminus S_i^0(x)$$

For each player, split the distinct online additions into canonical-trace and off-reference parts:

$$|U_i(x)| = |U_i(x) \cap C_i(x)| + |U_i(x) \setminus C_i(x)|$$

Summing over the two players yields

$$\sum_i |U_i(x)| \leq m_x + h_x$$

Each failed check produces one expansion and adds at least one distinct action, so $E_x$ is at most the total number of distinct online additions. This proves the expansion bound. The BR-call formula follows because the stated protocol invokes both players' full-space best responses once before any expansion and once after every expansion.

Why $h_x$ is necessary. The reference trace is only a canonical trajectory under its seed, tolerance, and tie-breaking. The online run can encounter off-reference best responses even when recall of canonical actions is high. Thus， recall reduces $m_x$, but it alone does not provide a universal bound without $h_x$.

**Supporting Checks and Sensitivity Analyses**

We next report supporting checks rather than additional primary comparisons. The T16 exact finite benchmark provides a fully enumerable reference setting in which exact exploitability and certificate gaps can be evaluated directly. As shown in following table.

| Variant | p95 Expl. | p95 Gap | False accept | Hidden g_D |
|---|---|---|---|---|
| Proposal-only (ridge) | 0.102 | 0.526 | -- | -- |
| Proposal-only (histGB) | 0.030 | 0.103 | -- | -- |
| Attacker-only (ridge) | 0.102 | 0.526 | 0.362 | 0.526 |
| Attacker-only (histGB) | 0.026 | 0.103 | 0.086 | 0.103 |
| Certified DO (learned-fast) | 0.010 | 0.042 | -- | -- |

***Note.*** *This compact table reports only the safety-critical ablations: proposal-only deployment, attacker-only certification, and full two-sided certified DO. It is not intended to replace the complete baseline comparison. "--" denotes metrics that are not applicable. The p95 values are computed on the returned outputs, while hidden $g_D$ is measured only among cases accepted by the attacker-only check.*

The certification diagnostics further show why the proposal cannot be deployed as a final strategy: proposal-only restricted solutions have substantially larger tail risk, while attacker-only checks can falsely accept solutions with large hidden defender-side gaps. These results are consistent with the CAP-DO design principle that learning proposes actions, whereas full-space certified Double Oracle controls acceptance. Increasing the proposal cap lowers p95 exploitability and expansion count, while tightening the certificate budget lowers the final gap at the cost of additional search. Learned-balanced remains stable across the tested cap and threshold settings. As shown in following table.

| Setting | p95 Expl. | p95 Gap | Mean Exp. | Time (s) |
|---|---|---|---|---|
| Proposal cap | | | | |
| cap = 10 | 0.024 | 0.047 | 1.896 | 0.104 |
| cap = 25 | 0.019 | 0.040 | 0.780 | 0.037 |
| cap = 50 | 0.011 | 0.032 | 0.158 | 0.024 |
| cap = 100 | 0.004 | 0.021 | 0.014 | 0.028 |
| Certificate threshold | | | | |
| tau = 0.10 | 0.017 | 0.062 | 0.066 | 0.021 |
| tau = 0.05 | 0.011 | 0.032 | 0.158 | 0.024 |
| tau = 0.02 | 0.004 | 0.016 | 0.300 | 0.028 |
| tau = 0.01 | 0.001 | 0.008 | 0.478 | 0.035 |

**Technical Supplement Table A1. Offline preparation cost**

| cost_type | mean_sec | model | sd_sec | stage |
|---|---|---|---|---|
| from-scratch preparation | 285.6329 | Both sides | 38.84715 | Reference trace generation |
| incremental learning | 0.0068 | Both sides | 0.004194 | Label assembly |
| incremental learning | 7.7257 | Both sides | 0.106414 | Training feature construction |
| incremental learning | 0.8722 | Ridge | 0.016498 | Ranker fitting |
| incremental learning | 5.1761 | HistGB | 0.826782 | Ranker fitting |
| incremental learning | 8.6047 | Ridge | 0.117621 | Incremental learning total |
| incremental learning | 12.9087 | HistGB | 0.719313 | Incremental learning total |
| from-scratch preparation | 294.2377 | Ridge | 38.77986 | From-scratch preparation total |
| from-scratch preparation | 298.5416 | HistGB | 39.27015 | From-scratch preparation total |

***Note.*** *T40 uses 100 training contexts per seed and 9,880 actions per player. Values are mean ± SD over three training seeds and include both defender and attacker rankers. Reference-trace generation is a one-time from-scratch cost; incremental learning assumes that reference traces are already available. From-scratch preparation total equals reference-trace generation plus incremental learning.*

**Technical Supplement Table A2 Per-context proposal overhead**

| model | mean_ms | sd_ms | stage |
|---|---|---|---|
| Shared | 75.0503 | 2.1138 | Feature construction |
| Ridge | 1.7565 | 0.3155 | Scoring + top-K (K=50) |
| Ridge | 76.8067 | 2.1442 | Total proposal overhead |
| HistGB | 45.7486 | 6.6676 | Scoring + top-K (K=50) |
| HistGB | 120.7989 | 7.0008 | Total proposal overhead |

***Note.*** *T40 contains 9,880 actions per player. Values are mean ± SD over 150 matched seed–context evaluations (three seeds × 50 test contexts) and include both defender and attacker rankers. Feature construction is shared by the two rankers. Total proposal overhead includes feature construction and scoring with top-K selection ($K = 50$), but excludes restricted-game solving, full-space best responses, certification, and subsequent DO iterations.*

**Technical Supplement Figure 1**

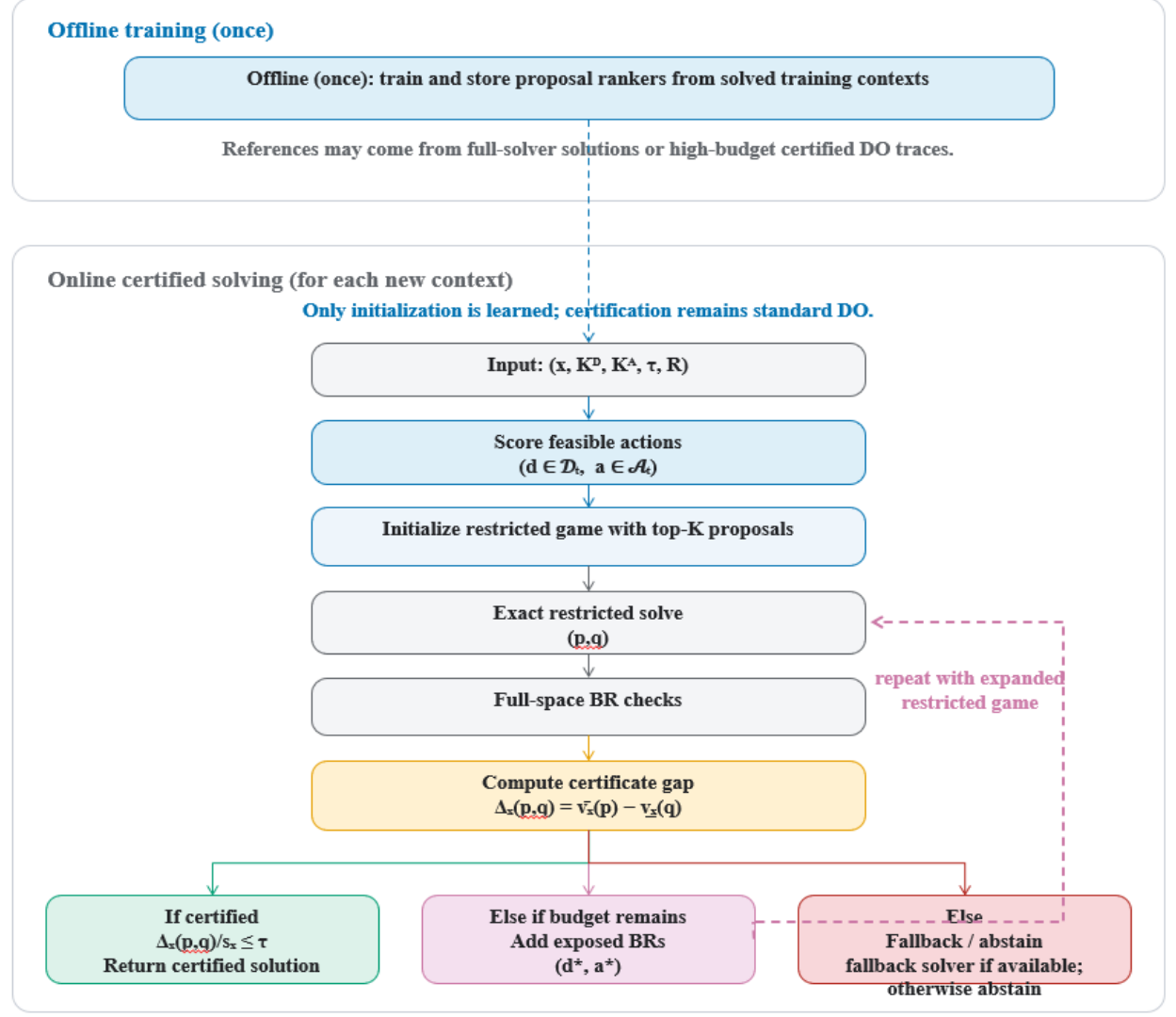

*Technical Supplement Figure 1: CAP-DO algorithm flow. Proposal rankers are trained once offline and reused for each new context. Online, CAP-DO initializes the restricted game with top-K proposed actions, solves the restricted game, computes full-space best-response bounds, and evaluates the two-sided certificate. Certified instances are accepted; otherwise, exposed best responses are added while the expansion budget remains. If certification fails after the budget is exhausted, the instance falls back to a full solver when available or abstains in oracle-only regimes.*